\documentstyle[sprocl]{article}
\newcommand{\be}{\begin{equation}}
\newcommand{\ee}{\end{equation}}
\newcommand{\bea}{\begin{eqnarray}}
\newcommand{\eea}{\end{eqnarray}}

\newcommand{\bal}{\begin{array}{ll}}
\newcommand{\eal}{\end{array}}

\def\thru#1{\mathrel{\mathop{#1\!\!\!\!/}}}
\def\thru#1{\mathrel{\mathop{#1\!\!\!\!/}}}
\def\1{{\rm 1 \kern -.10cm I \kern .14cm}} \def\R{{\rm R \kern -.28cm I
\kern .19cm}}
\begin{document}
%\begin{titlepage}
%\begin{flushright} UFIFT-HEP-97-17\\ RU-97-32
%\begin{flushright}%August 23, 1999%\end{flushright}%\vskip .8cm

\title{DIRAC EQUATIONS\\ LIGHT CONE SUPERSYMMETRY\\   and\\ SUPERCONFORMAL ALGEBRAS}
\author{Lars Brink}
\address{ Department of Theoretical Physics,\\ Chalmers University
of Technology,\\  S-412 96 G\"oteborg, Sweden}
\author {Pierre Ramond}
\address{ Institute for
Fundamental Theory, Department of Physics,\\
University of Florida,  Gainesville FL 32611, USA}
\maketitle\abstracts{
 After a brief historical survey that emphasizes the role of the algebra
obeyed by the Dirac
operator,  we examine an algebraic Dirac operator associated with Lie
algebras and  Lie algebra
cosets. For symmetric cosets, its ``massless'' solutions  display
non-relativistic supersymmetry, and
can be identified with the massless degrees of freedom of some
supersymmetric theories: $N=1$
supergravity in eleven dimensions (M-theory), type IIB string theory in ten and
four dimensions, and in four dimensions, $N=8$ supergravity, $N=4$
super-Yang-Mills, and the
$N=1$ Wess-Zumino multiplet.  By generalizing this
Dirac operator to the affine case, we generate superconformal algebras
associated with cosets
${\bf g}/\bf h$, where
$\bf h$ contains the {\it space} little group. Only for eleven
dimensional supergravity is $\bf h$ simple. This suggests, albeit
in a non-relativistic setting, that these may be the limit of theories with 
underlying two-dimensional superconformal structure.}

\section{Historical Remarks}
The development of supersymmetry in the Western world and in the Eastern
world of the time was very different. In this book there are a number of
contributions that tell the story from the Eastern side. In the Soviet
scientific society the scientists had one freedom that scientists in the
West lacked and still lack (perhaps the only real freedom that Eastern
scientists had),
and that was to spend time also on esoteric questions. They did not have to
be scrutinized by funding agencies every now and then. As one consequence of
this fact, the Soviet scientists of the time were often very well trained in
basic formalisms. They had in many cases a superior knowledge in fermion
fields and their treatment in quantum field theory, which is very well
witnessed in the brilliant book by Berezin~\cite{Ber}. This background was
fertile
ground for work leading up to supersymmetry. To us it seems that the
development resulting in the work of Yurij Golfand and Evgenij
Likhtman~\cite{Gol} was
a natural one (without meaning to underestimate its fundamental importance).
By studying quantum field theory per se without the need to directly
correlate it to particle physics of the time it is a clever but also a
natural idea to look for a symmetry between  bosons and  fermions.
Quantum field theory that was essentially abandoned in the West because of
its severe difficulties to go beyond QED, certainly needed new elements. To
match the number of bosonic degrees of freedom to the fermionic ones was not
a natural idea in the West in the 60's which was in the midst of the Quark
Model and Current Algebra. Of course, a superalgebra in terms of the
anticommutator of the two Dirac operators was certainly known but not given
much importance. Abdus Salam and collaborators~\cite{Abdus} had been
interested in
functional integrations over spinors in the 50's but had not completed the
issue as Berezin had done. However, once four-dimensional supersymmetry
became popular in the West, Abdus Salam became very active and had a huge
experience to draw from. However, Yurij Golfand might very well have been
the only one who seriously pushed a four-dimensional fermionic symmetry in
quantum field
theory in the late 60's.

In the West the abandonment of quantum field theory for the strong
interactions led to S-matrix theory and to a detailed study of the possible
analytic structure of this matrix. More and more resonances were found in
the strong interactions experiments, and in 1968 Gabriele
Veneziano~\cite{Gab} managed
to construct an S-matrix (a dual resonance model) for four bosons that
incorporated infinitely many resonances with the necessary analytic
structure for a Born-term. This was the starting point for string
theory in that slightly later it was understood to describe scattering of
string states. The important point we want to make is that this development
was very much influenced by the experimental discoveries of the time.
Particle physics was still quite  young and the theorists in the West were
very much involved in constructing models for the experimental data that
kept pouring out. After Veneziano's fundamental work it was so natural to
ask how to implement fermions into dual resonance model. This was not an
easy task since the model was formulated just as a scattering amplitude. The
key to solve this was the realization of the underlying two-dimensional
world sheet. Veneziano's model described scattering based on bosonic degrees
of freedom on the world-sheet. To introduce space-time spinors the natural
thing was then to introduce fermions on the world sheet~\cite{Frog}. In this
process it was realized that two-dimensional supersymmetry was needed to get
a viable model and this was the start of supersymmetry in the West. The
symmetry even had to be a superconformal symmetry and the ordinary
supersymmetry came out as a zero-mode. Supersymmetry hence was a result of the
direct strive to find a model to describe the strong interactions. The fact
that the two first papers of supersymmetry in the West and the East were
submitted
at about the same time in late 1970 is quite remarkable.

In the 30 years since this development took place the two lines of
theoretical research described above have converged. The development of
supersymmetric gauge field theories and superstring theory took us out of
contact with much of the experimental data of today. The success of
superstring theory builds to a large extent on our increased mathematical
insight and this has spurred us to look for new structures. The science of
today is very much depending on our attempts to find those and now we cannot
only rely on experimental data to lead our way. The studies of superstring
theory
and supersymmetric gauge theory per se have led us to all the wonderful
discoveries of recent years. It is hence very much worthwhile to study new
structures in mathematics and try to connect such developments to modern
string theory. The understanding of M-theory as an 11-dimensional theory
which is not a regular string theory has made us go back to a study of group
theory to look for alternatives to supersymmetry and in this contribution we
will discuss some recent studies of ours which we hope will shed new light
on even more fundamental levels of physics than superstrings.

\section{Supersymmetry and the Dirac Equation}
It was known since the early days of quantum mechanics that the Dirac equation
\be
\thru p\Psi=0\ ,
\ee
contains a new type of anticommuting symmetry, since the Dirac operator
obeys the superalgebra
\be
\{~\thru p,\thru p~\}=p^2\ ,\qquad [~\thru p,p^2~]=[~p^2,p^2~]=0\ .
\ee
However, in ordinary quantum field theory this symmetry was not really
used. In the development of
the Veneziano Model in the late 60's it was however realized that the modes of
$p^2$ when taken to be a function of a complex variable (of the
Euclideanized world-sheet) span the
two-dimensional conformal algebra, the Virasoro algebra~\cite{Mig}. This
algebra is
the underlying symmetry
algebra of the model and it was natural to seek to generalize this algebra
to also contain
fermions~\cite{Frog}. The key then was to generalize the $\gamma$-matrices to also be
functions of this complex
variable. This led to the superconformal algebra and eventually to the
superstring model. As a
by-product one got the 2-dimensional supersymmetry algebra. Strings
naturally demanded
2-dimensional supersymmetry and eventually it was also realized that they
demand supersymmetry in
the target space.

In the struggle to find realistic string models the first attempts were to
generalize the
superconformal algebra. Such a generalization was finally found in the
$N=2$ string~\cite{football} where $N$
refers to the underlying superconformal algebra. When the superstring was
properly understood as
unifying theories of all interactions it was clear that the original one
was the most promising and
the challenge was then to understand the 10-dimensional space on which it
lives. In this way we
obtained the five superstring models and opened up the road to
compactifications. With the advent
of M-theory as a non-standard string theory we have found it worthwhile to study
again the roots of
supersymmetry and its role in string theory. The superconformal generator
is part of the
superconformal algebra used as the gauge algebra of the model and since in
string theory the target
manifold contains the four-dimensional Minkowski space the zero mode part
gives that states in the
R-sector must satisfy the Dirac equation in Minkowski space. In the sequel
we will study Dirac
equations on other types of manifolds and check when the ``massless" states
coincide with
interesting multiplets.

\section{Kostant's Operator}
As we have seen, in formulating supersymmetric theories, it is natural to
focus on the Dirac operator. It is then interesting to note that for any
manifold with
dimension equal to that of a Lie algebra, or for any coset generated Lie
algebra embeddings, there
exists a natural Dirac-like operator~\cite{KOS}.  For compact Lie algebras,
this Dirac operator has
only ``massive'' solutions. However, the Dirac operator associated with
symmetric cosets has
``massless'' solutions, some of which display manifest supersymmetry. In
the sequel we will use
this method to generate supermultiplets. The method will be group
theoretical and the relation to
space-time symmetries will not be obvious.

\subsection{ Lie Algebras}
Consider a $D$ dimensional Lie algebra, $\bf g$, defined by the commutation
relations
\be
[T_{}^a,T_{}^b]=if_{}^{abc}T_{}^c\ , \ee
with $a,b,c=,\dots, D$. On this $D$-dimensional vector space, the $D$ Dirac
matrices which satisfy the Clifford algebra
\be
\{\gamma^a,\gamma^b\}=2\delta^{ab}\ ,\ee
are realized in terms of $2^D\times 2^D$ matrices (for $D$ even). We
now define the Kostant-Dirac operator

\be\thru {\cal K}~\equiv \sum_{a=1}^D\gamma_{}^a T_{}^a-{i\over 2\cdot
3!}\sum_{a,b,c}^{D}
\gamma_{}^{abc} f_{}^{abc}\ ,\ee
where $\gamma^{abc}=\gamma^{[a}\gamma^b\gamma^{c]}$ is totally
antisymmetrized. Because of the term cubic in the Clifford algebra, it has
the remarkable property
that
\be
\{\thru {\cal K},\thru {\cal K}\}~=~2C_2+{D\over 12}C_2^{\rm adjt}\ ,\ee
where $C_2$ is the quadratic Casimir operator of the representation of the
$T_{}^a$ matrices,
 and $C_2^{\rm adjt}$ is the quadratic Casimir of the adjoint
representation of the algebra.
Because the algebra is compact, the square is the sum of two positive
terms, and the solution to the $``$massless" equation $\thru {\cal K}~\Psi=0$ is trivial, 
although there are many states on which $\thru {\cal K}$ takes on a
constant value. We note that
\be
[~L_{}^a,\thru {\cal K}~]=0\ ,\ee
where the diagonal  matrix operators
\be
L_{}^a=T_{}^a+S_{}^a\ ,\ee
with
\be
S_{}^a=-{i\over 4}\gamma_{}^{bc} f_{}^{abc}\ ,\ee
generate the representation of the algebra $\bf g$ on the spinor space of
$SO(D)$. Thus the
solutions of $(\thru {\cal K}-\mu)\Psi=0$ will assemble in full multiplets
of $\bf g$ for special
values of the $``$mass" $\mu$.

\subsection{Lie Algebra Cosets}
Consider any Lie algebra, $\bf g$, generated by the $T_{}^a$,
$a=1,2,\dots,D$. Inside that Lie algebra lies a subalgebra, ${\bf  h}\subset 
{\bf g}$, generated by $T_{}^i$, $i=1,2,\dots,d$. We label the
remaining $(D-d)$
generators as $T_{}^m$. The commutation relations are
\bea
[~T_{}^{i~},T_{}^{j~}]&=&if_{}^{ijk}T_{}^k\ , \cr
[~T_{}^{i~},T_{}^{m~}]&=&if_{}^{imn}T_{}^n\ , \cr
[~T_{}^m,T_{}^n]&=&if_{}^{mnj}T_{}^j+if_{}^{mnp}T_{}^p\ .
\eea
Consider the $(D-d)$ even-dimensional coset space, where the $\gamma^m_{}$
are $2^{D-d}\times 2^{D-d}$ matrices that satisfy
\be
\{\gamma^m,\gamma^n\}=2\delta^{mn}\ .\ee

The Kostant operator associated with the coset is defined as
\be\thru {\cal K}~\equiv \sum_{m=1}^{D-d}\gamma_{}^m T_{}^m-{i\over 2\cdot 3!}
\sum_{m,n,p}^{D-d}\gamma_{}^{mnp} f_{}^{mnp}\ .\ee
 Simple algebra and use of the commutation relations yield

\be
\thru {\cal K}^{~2}~=~\Bigl[C_2^{}({\bf g})+{D\over 24}C_2^{\rm
adjt}({\bf g})\Bigr]-
\Bigl[C_2^{}({\bf h})+{d\over 24}C_2^{\rm adjt}({\bf h})\Bigr]\ ,\ee
which is not positive definite:   the ``massless'' Kostant
equation can have  non-trivial solutions. The operators
\be
S_{}^i=-{i\over 4}\gamma_{}^{mn} f_{}^{imn}\ ,\ee
represent the subalgebra $\bf h$ on the $(D-d)$ coset space. The subalgebra
$\bf h$ is generated by the diagonal sum
\be
L_{}^i=T_{}^i+S_{}^i\ ,{\rm ~~~i=1,2\dots,d}\ee
which commute with the Kostant operator
\be
[L_{}^i~,\thru {\cal K}~]=0\ .\ee
Hence, solutions of
\be
\thru {\cal K}~\Psi=~0\ ,\ee
can be assembled in full multiplets of the subalgebra $\bf h$.

When  $\bf g$ and $\bf h$ have the same rank, these
solutions  reside in the tensor product~\cite{GKRS,LERCHE}
\be
V_\lambda\otimes S^+ - V_\lambda\otimes S^-\ ,\label{eq:diff}\ee
where  $S^\pm$ are the two spinor representations of $SO(D-d)$,  and
$V_\lambda$ is a representation of $\bf g$ of highest weight $\lambda$, with
all  representations decomposed in terms of those of the subalgebra $\bf h$.

We specialize our discussion to  {\em symmetric} coset space $\bf
{g/h}$, for which  $f^{}_{mnp}=0$~\footnote{when the coset space is not
symmetric, $f^{}_{mnp}\ne 0$,
there
are always cancellations, even when $V_\lambda$ is a singlet, and no manifest
supersymmetry.}.  In that case, it is only when $V_\lambda$ is the
singlet of $\bf g$ that there are no
cancellations of representations in eq.(\ref{eq:diff}). It follows that the solutions 
of the Kostant-Dirac equation
span the two spinor representations of $SO(D-d)$. Since these are well-known to be
generated by a Clifford algebra, they necessarily exhibit 
supersymmetry in terms of the subalgebra $\bf h$.

On the other hand, if  $V_\lambda$ is taken to be any non-trivial
representation of $\bf g$,  cancellations always occur, and the solutions no
longer display manifest supersymmetry.

It is also known that the number of irreps of $\bf h$ appearing in this
difference is always
the same, and equal to $r$, the ratio of the orders of the Weyl groups
of $\bf g$ and $\bf h$, which is also the Euler number of the coset
manifold.  In some cases, the r-plets, or Euler multiplets,  have the same number of bosons and
fermions~\cite{PR}, but no apparent supersymmetry, which
make them very interesting from a quantum mechanical point of view. We will
not pursue this line
of thinking in this paper. Indeed, it has been shown that
\be
V_\lambda\otimes S^+ - V_\lambda\otimes S^-=\sum_{c}^r (-1)^w~U^{}_{c\bullet
\lambda}\ ,\ee
where the sum is over the $r$ elements of the Weyl group of $\bf g$ that
are not in $\bf h$'s, and $U_{c\bullet
\lambda}$ denote representations of $\bf h$, and the $\bullet$ denotes a 
specific construction which has been defined elsewhere~\cite{GKRS,LERCHE}.

To summarize, the solutions of the massless Kostant operator  appear in
Euler multiplets, one for each representation of $\bf g$. Each Euler
multiplet contains exactly  $r$ irreps of the subalgebra $\bf h$.
For symmetric coset spaces, the Euler multiplet associated with
the singlet of $\bf g$ spans a Clifford algebra, and therefore displays
supersymmetry.

\subsection{Supersymmetric Euler Multiplets}
It is well-known that the degrees of freedom of supersymmetric
theories can be labelled in terms of the Wigner little group of the
associated Poincar\'e algebra. Since theories with gravity cannot sustain a
finite number of {\it massless} degrees of freedom with spin higher than two, the cosets
cannot exceed sixteen dimensions, which yield the two  spinor representations of $SO(16)$,
 with 256 degrees of freedom.

We have found only a few cosets, with $D-d=16,8,4$, for which the basic
Euler multiplets have the right quantum numbers to represent massless particles of relativistic theories:

\vskip .5cm
\noindent $\triangleright$ {\bf $16$-dimensional Cosets}
\vskip .3cm
\noindent The $256$ states of the associated Clifford  are generated by the
two spinor irreps of
$SO(16)$, yielding four possibilities:
\begin{itemize}
\item $SU(9)\supset SU(8)\times U(1)$
with lowest Euler multiplet
$$ {\bf 1}_2 \oplus {\bf 8}_{3/2}\oplus  {\bf 28}_1 \oplus {\bf 56
}_{1/2}\oplus {\bf 70}_0 \oplus
{\overline {\bf 56}}_{-1/2} \oplus {\overline {\bf 28}}_{-1}\oplus
{\overline {\bf 8}}_{-3/2}
\oplus {\bf 1}_{-2} $$ Interpreting the $U(1)$ as the helicity little group
in four dimensions,
these can be thought of as  the massless spectrum of either type IIB string
theory,  or  of
$N=8$ supergravity.
\item $SO(10)\supset SO(8)\times SO(2)$
with lowest Euler multiplet
$$ {\bf 1}_{2}\oplus  {\bf 8}_{3/2}\oplus  {\bf 28}_{1} \oplus {\bf
56}_{1/2} \oplus ({\bf 35}_{0}
\oplus {\bf 35}_{0}) \oplus {\bf 56}_{-1/2} \oplus {\bf 28}_{-1}\oplus
{\bf 8}_{-3/2}\oplus  {\bf
1}_{-2} $$

Viewing $SO(8)$ as the little group in ten dimensions, these may represent
the massless spectrum of
IIB superstring in ten dimensions, but only after using $SO(8)$ triality.
Alternatively, with
$U(1)$ as the helicity, it could just be the massless particle content of the
type IIB superstring
theory, or of $N=8$ supergravity in four dimensions.
\item $SU(6)\supset SO(6)\times SO(3)\times SO(2)$
with lowest Euler multiplet
$$({\bf 1},{\bf 1})_{2}\oplus ({\bf 4},{\bf 2})_{3/2}\oplus ({\bf
10},{\bf 1})_1
\oplus ({\bf 6},{\bf 3})_1\oplus ({\bf 4},{\bf 4})_{1/2}\oplus ({\bf
20},{\bf 2})_{1/2}$$
$$~~~~~~~~~~~\oplus ({\bf 20^\prime},{\bf 1})_0\oplus ({\bf 15},{\bf 3})_0
\oplus ({\bf 1},{\bf 5})_0$$
$$
\oplus ({\bf 4},{\bf 4})_{-1/2}\oplus ({\overline{\bf 20}},{\bf 2})_{-1/2}
\oplus ({\overline{\bf
10}},{\bf 1})_{-1}
\oplus ({\bf 6},{\bf 3})_{-1}\oplus ({\overline{\bf 4}},{\bf 2})_{-3/2}
\oplus({\bf 1},{\bf 1})_{-2}\ .$$
This is the massless spectrum of $N=8$ supergravity, or of type IIB superstring
in four dimensions, or of massless theories in five and eight dimensions.

\item $F_4\supset SO(9)$
The lowest Euler multiplet
$$ {\bf 44}\oplus  {\bf 84}\oplus  {\bf 128} $$
describes the massless spectrum of $N=1$ supergravity in $11$ dimensions,
the local limit of M-theory, by identifying $SO(9)$ as the massless little
group.

\end{itemize}
\vskip .3cm
\noindent $\triangleright$ {\bf $8$-dimensional Cosets}
\vskip .3cm
\noindent The Clifford algebra is realized on the $16$ states that span the
two spinor irreps of
 $SO(8)$, leading to the two massless interpretations
\begin{itemize}
\item $SU(5)\supset SU(4)\times U(1)$
 which yields the  lowest Euler multiplet
$$ {\bf 1}_1\oplus {\bf 4}_{1/2} \oplus  {\bf 6}_0 \oplus  {\overline {\bf
4}}_{-1/2} \oplus
{\bf 1}_{-1}$$
With $U(1)$ as the helicity little group, this particle content is the same
as the massless
spectrum of $N=4$ Yang-Mills in four dimensions.

It could also describe a massless theory in eight dimensions with one
conjugate spinor, one vector, and two
scalars, since $SU(4)\sim SO(6)$.

\item $SO(6)\supset SO(4)\times SO(2)$
with lowest Euler multiplet
$$ ({\bf 1},{\bf 1})_1\oplus  ({\bf 2},{\bf 2})_{1/2} \oplus ({\bf 1},{\bf
3})_0\oplus  ({\bf 3},
{\bf 1})_0\oplus  ({\bf 2},{\bf 2})_{-1/2}\oplus  ({\bf 1},{\bf 1})_{-1}$$
This particle content is the same as the $N=4$ Yang-Mills in four
dimensions, or the $N=2$ vector supermultiplet in five dimensions.

\end{itemize}
\vskip .3cm

\vskip .3cm
\noindent $\triangleright$ {\bf $4$-dimensional Cosets}
\vskip .3cm
\noindent The Clifford algebra is realized on $4$ states of the two spinor
irreps of $SO(4)$.
They appear as the lowest Euler multiplet in the following decomposition:
\begin{itemize}
\item $SU(3)\supset SU(2)\times U(1)$.
The lowest Euler multiplet$$ {\bf 1}_{1/2}\oplus {\bf 2}_{0}\oplus {\bf
1}_{-1/2}$$
can be interpreted as a relativistic theory in $4$ dimensions, since the
$U(1)$ charges are half
integers and integers, and it can describe the massless $N=1$ Wess-Zumino
supermultiplet in four
dimensions, as well as a massless supermultiplet in five dimensions.
\end{itemize}

So far, we have limited our discussion to massless degrees of freedom, but
we should note for
completeness that possible relativistic theories with  massive degrees of
freedom appear in the
cosets $Sp(2P+2)\supset Sp(2P)\times Sp(2)$ with the same content as
massive $N=P$ supersymmetry in
four dimensions. The massive $N=4$ supermultiplet in four dimensions is
also described by the coset
$SO(8)\supset SO(4)\times SO(4)$, and the massive $N=2$ supermultiplet in
four dimensions can be described in terms of $G_2/SU(2)\times SU(2)$.

In this section, we have shown that the degrees of freedom of some
supersymmetric theories are solutions of the Kostant-Dirac
equation associated with specific  cosets. While non-relativistic, this
identification is suggestive of an alternate description, in which the
states appear as solutions of an operator
equation, not as fundamental fields in a Lagrangian. This is of course
reminiscent of what happens in string theory. 
Furthermore, not all supersymmetric theories appear in this
list, only the local limit of M-theory, of type
IIB superstring theories (not type IIA, type I,
nor heterotic superstrings), and certain local field theories, $N=4$
Yang-Mills, and $N=1$ Wess-Zumino multiplets. All the massless cosets
are both hermitian and symmetric, except for the Wess-Zumino multiplet
and $N=1$ supergravity in eleven dimensions which are only symmetric.

To take advantage of these alternate descriptions, the Kostant-Dirac
operator must be given a fundamental role. In the next section,  we
propose one approach where  the Kostant-Dirac operator is the zero
mode of the superconformal generator associated with the coset,
constructed some years ago by Kazama and Suzuki~\cite{KS}.

\section{Coset Construction of Superconformal Algebras}
One can construct an affine generalization of the Kostant-Dirac operator,
much in the same way the
Dirac equation was generalized many years ago: one generalizes the Clifford
matrices to the world sheet as
\be
\gamma^a\to \Gamma^a(\sigma)\ ,\ee
with
\be
\{\Gamma^a(\sigma),\Gamma^b(\sigma')\}=2\delta^{ab}\delta(\sigma-\sigma')\ ,\ee
as well as the generators
\be
T^a\to T^a(\sigma)\ ,\ee
which satisfy the level $k$ Kac-Moody commutation relations
\be
[T^a(\sigma),T^b(\sigma')]=if^{abi}T^i(\sigma)\delta(\sigma-\sigma')+k\delta
^{ab}
\delta^\prime(\sigma-\sigma')\ .\ee
 These are the cosets of the full Kac-Moody
algebra of affine $\bf
\hat g$, with the additional commutation relations (we limit our discussion
to symmetric cosets)
\be
[T^i(\sigma),T^b(\sigma')]=if^{iab}T^b(\sigma)\delta(\sigma-\sigma')\ ,\ee
\be
[T^i(\sigma),T^j(\sigma')]=if^{ijk}T^k(\sigma)\delta(\sigma-\sigma')+k\delta
^{ij}
\delta^\prime(\sigma-\sigma')\ .\ee
These two algebras are independent and commute with one another
\be
[\Gamma^a_{}(\sigma),T^b_{}(\sigma')]=0\ .\ee
The generators
\be
F(\sigma)\equiv\Gamma^a(\sigma)T^a(\sigma)=\sum_{n=-\infty}^{+\infty}F^{}_ne
^{-in\sigma}\ ,\ee
generate a superconformal algebra
\be \{F^{}_n,F^{}_m\}=2L^{}_{n+m}+{c\over 3}(n^2-{1\over
4})\delta_{m+n,0}\ ,\ee
where the $L_n$ generate the Virasoro algebra
\be [L_n,L_m]=(m-n)L_{m+n}+{c\over 12}m(m^2-1)\delta_{m+n,0}\ ,\ee
and
\be
[L_n,F_m]=({n\over 2}-m)F_{m+n}\ .\ee
Alternatively, the Virasoro generators can be summarized in
\be
L(\sigma)=T_{}^a(\sigma)T_{}^a(\sigma)+if_{}^{abj}\Gamma_{}^a(\sigma)
\Gamma_{}^b(\sigma)T_{}^j(\sigma)-
k\Gamma_{}^a(\sigma){d\Gamma^a(\sigma)\over d\sigma}\ ,\ee
after getting rid of a surface term. This can be rewritten as
\be
L(\sigma)=L^{\bf g}_{}(\sigma)-L^{\bf h}_{}(\sigma)- {1\over 4}
f_{}^{abj}f_{}^{cdj}\Gamma_{}^a(\sigma)
\Gamma_{}^b(\sigma)\Gamma_{}^c(\sigma)
\Gamma_{}^d(\sigma)\ ,\ee
in which
\be
L^{\bf g}_{}(\sigma)=T_{}^A(\sigma)T_{}^A(\sigma))-
k\Gamma_{}^A(\sigma){d\Gamma^A(\sigma)\over d\sigma}\ ,\ee
\be
L^{\bf h}_{}(\sigma)=\hat T_{}^j(\sigma)\hat T_{}^j(\sigma))-
k\Gamma_{}^i(\sigma){d\Gamma^i(\sigma)\over d\sigma}\ ,\ee
where
\be
\hat T_{}^j(\sigma)=T_{}^j(\sigma) -{i\over 2}f_{}^{abj}\Gamma_{}^a(\sigma)
\Gamma_{}^b(\sigma)\ ,\ee
generate the diagonal Kac-Moody associated with the subalgebra $\bf
h$. This is the construction of Kazama and Suzuki~\cite{KS}, for which  we have
\be
c={3\over 2}{\rm dim}({\bf g}/{\bf h}){k\over k+g}\ ,\ee
where $g$ is the second Dynkin index of the adjoint representation of $\bf
 g$ (dual Coxeter number), and $k$ is the level  of the $\bf\hat g$
Kac-Moody
algebra. 

It is convenient to introduce the expansions
\be
\Gamma^a(\sigma)=\gamma^a+\sqrt{2}\gamma_5\sum_{n\ne 0}b_n^ae^{-in\sigma}\ ,\ee
where
\be
\{\gamma^a,\gamma_5\}=0\ ,\qquad
\{b^a_n,b^b_m\}=\delta^{ab}_{}\delta^{}_{m+n,0}\ ,\ee
and
\be
T^a_{}(\sigma)=T^a_{}+\sum_{n\ne 0} T^a_ne^{-in\sigma}\ ,\ee
with
\be
[T^a_n,T^b_m]=if_{}^{abj}T^j_{m+n}+km\delta^{ab}_{}\delta^{}_{m+n,0}\
.\ee
We then see that
\be
F^{}_0=\thru{\cal K}+~\sqrt{2}\gamma_5\sum_{n\ne 0} b^a_nT^a_{-n}\ ,\ee
is the natural generalization of the Kostant operator. The oscillator ground states satisfy
\be
b_n^a\vert\Psi_0>=0\ ,\qquad T_n^a\vert \Psi_0>=0\ ;\qquad n>0\ .\ee
Hence, the  massless degrees of freedom of the previous section, are the
ground states of the
superconformal algebra in the fermion sector. These states form Euler
multiplets of representations
of $\bf h$,  but they contain no oscillator modes.

The oscillator modes of this superconformal algebra come from two places,
the $b^a_n$ which appear
in the generalized Dirac matrices, and transform as spinors  of $\bf
h$,  and the modes
which appear in the construction of the $\bf\hat g$ Kac-Moody algebra. There are
several ways to generate this algebra~\cite{GO}, by means of:
\begin{itemize}
\item  a current algebra as a bilinear in fermion oscillators transforming as
some representation of
 $\bf g$, in which case, $k$ is  the second-order Dynkin index of the representation.
\item  a bosonic WZNW construction.
\item  a vertex construction, only if $\bf g$ is simply-laced, with
$k=1$.
\end{itemize}
Each of these constructions yields a definite level $k$, and therefore
a different $c$-number for the associated superconformal algebra.

The role of the $c$-number in this work is not yet clear, since the use of
the superconformal
algebra advocated here is hardly conventional: the unbroken subgroup $\bf
h$, or parts thereof is
not an internal symmetry group,  but a {\em space} group. As a consequence
{\em both space-time bosons and
fermions reside in the R-sector}. On the other hand, the description is
non-relativistic: the Hilbert space is manifestly positive definite, 
but relativistic invariance is highly non-trivial to prove, and we 
surmise that relativistic invariance could be possible, if at all, 
only for special values of the anomaly.

In this work, we do not attempt a proof of relativistic invariance, but
only offer two examples which set forth {\em necessary} conditions 
for these theories to be relativistic.

Consider the lowest Euler multiplet of the $F_4/SO(9)$ coset theory, which  describes the
massless degrees of freedom of the $N=1$ Supergravity in eleven dimensions, with $SO(9)$
identified as the massless little group.  
The oscillators of this theory come from two places, those in the 
generalized Dirac matrices, which  transforms as the real $\bf 16$ spinor of $SO(9)$,
and the oscillators which build up the $F_4$ Kac-Moody algebra.

If the oscillator modes of this superconformal theory are to describe 
massive degrees of freedom, a necessary condition for a relativistic description is 
that they assemble themselves in representations of the {\em massive} little group, $SO(10)$. As we s
ee in the two examples below, this depends on the realization of the Kac-Moody algebra:

\begin{itemize}
\item
First, we assume a current algebra realization of the $F_4$ Kac-Moody algebra, with 
the fermions transforming as the lowest dimensional representation, the
real $\bf 26$. In terms of
$SO(9)$,  ${\bf 26}={\bf 1}\oplus{\bf 9}\oplus{\bf 16}$.
 Thus the Hilbert space of this superconformal coset algebra is generated by

{\obeylines
-- two oscillators, each transforming as the real {\bf 16} of $SO(9)$,
-- one $SO(9)$ vector oscillator
-- one $SO(9)$ singlet oscillator.}

These degrees of freedom could be assembled in representations of
$SO(10)$: the two real spinors into the
{\em complex} $\bf 16$
spinor of $SO(10)$, and the singlet and $SO(9)$ vector
into the vector $\bf 10$ of $SO(10)$. This counting does not prove that
this theory can be made relativistic, only that some necessary conditions are met.

\item As a second example, consider a current algebra with fermions in the
$\bf 52$ of $F_4$.
Since ${\bf 52}={\bf 36}\oplus {\bf 16}$,  we end  up with 

{\obeylines
-- two oscillators, each transforming as the real {\bf 16} of $SO(9)$,
-- one  oscillator transforming as the $\bf 36$.}

To obtain a representation of $SO(10)$, we would need  
an $SO(9)$ vector in  to complete the $\bf 45$ of
$SO(10)$.  Thus a relativistic description seems  implausible in this case.
\end{itemize}
Those two cases are distinguished by their  Kac-Moody level,  and thus
by the $c$-number of the consequent Virasoro algebra. 
One value may lead to a relativistic theory, while another cannot.

Finally, we note the values of the conformal anomaly for the  cosets
we have listed, suggesting possible connections between
(massless and massive) theories:
\begin{itemize}
\item $c=24k/(k+9)$ for $SU(9)/SU(8)\times U(1)$ and $F_4/SO(9)$.
\item $c=24k/(k+8)$ for $SO(10)/SO(8)\times U(1)$.
\item $c=24k/(k+6)$ for $Sp(10)/Sp(8)\times Sp(2)$.
\item $c=12k/(k+5)$ for $SU(5)/SU(4)\times U(1)$.
\item $c=12k/(k+4)$ for $SO(6)/SO(4)\times SO(2)$,  $Sp(6)/Sp(4)\times
Sp(2)$, and $G_2/SU(2)\times SU(2)$.
\item $c=6k/(k+3)$ for $SU(3)/SU(2)\times U(1)$.
\end{itemize}
Work is in progress to determine the representations  of these superconformal
algebras, in order to see if their spectrum displays relativistic properties.

\section{Euler Multiplets for $SU(3)/SU(2)\times U(1)$}
We know from the work of Lerche et al~\cite{LERCHE},
Kostant~\cite{KOS} and GKRS~\cite{GKRS} that for this equal rank
embedding, there are an infinite number of  Euler triples. Associated
with the $[a_1^{}~a_2^{}]$ representation of $SU(3)$ (in
Dynkinese), the solutions consist of the $SU(2)\times U(1)$ multiplets
\be
[a_1^{}+a_2^{}+1]_{(a_1^{}-a_2^{})/6}\oplus [a_1^{}]_{(2a_2^{}+a_1^{}+3)/6}
\oplus [a_2^{}]_{-(2a_1^{}+a_2^{}+3)/ 6}\ .\ee
It is instructive to show in detail how these solutions arise.

It is convenient to act on the  four-dimensional coset with the $4\times 4$
matrices
\be
\gamma^{}_4=\sigma_1\times \sigma_1\ ;\qquad
\gamma^{}_5=\sigma_1\times \sigma_2\ ;\qquad
\gamma^{}_6=\sigma_1\times \sigma_3\ ;\qquad
\gamma^{}_7=\sigma_2\times 1\ ,
\ee
which satisfy the Clifford algebra
\be
\{\gamma_{}^a,\gamma_{}^b\}=2\delta_{}^{ab}\ ,\ee
for $a,b=4,5,6,7$. The representation of the $SU(2)\times U(1)$ subgroup on
the spinors is given by
\be
S_{}^j=-{i\over 4}\gamma_{}^{ab}f_{}^{abj}\ ,\qquad S_{}^8=-{i\over
4}\gamma_{}^{ab}f_{}^{8ab}\ ,\ee
where $j=1,2,3$, and  $f_{}^{jab}$ and $f_{}^{8ab}$ are the antisymmetric
$SU(3)$ structure
functions which appear in

\be [T_{}^A,T_{}^B]=if_{}^{ABC}T_{}^C\ ,\ee
with $A,B,C=1,2,\dots, 8$, with   $f^{123}=1$, and other nonzero values
\be
f_{}^{147}=-f_{}^{156}=f_{}^{246}=f_{}^{257}=f_{}^{345}=-f_{}^{367}={1\over
2}\ ;
~~~f_{}^{845}=f_{}^{867}={\sqrt{3}\over 2}\ .\ee
%We find the explicit representations
%\bea
%S_{}^1&=&~~{1\over 4}(\sigma_3-1)\times \sigma_1\ ,\qquad
%S_{}^2={1\over 4}(\sigma_3-1)\times \sigma_2\ ,\\
%S_{}^3&=&-{1\over 4}(\sigma_3-1)\times \sigma_3\ ,\qquad
% S_{}^8={\sqrt{3}\over 16}(\sigma_3+1)\times \sigma_3\ .\eea
The Kostant operator is given by
\be\thru {\cal K}~\equiv \sum_{a=4}^7\gamma_{}^aT_{}^a\ ,\ee
 and is invariant under the action of the diagonal generators of the
$SU(2)\times U(1)$
\be
L_{}^j=T_{}^j+S_{}^j\ ,\qquad L_{}^8=T_{}^8+S_{}^8\ .\ee
To solve the Kostant-Dirac equation

\be\thru {\cal K}~\Psi= 0\ ,\ee
introduce
\be
\Psi=\pmatrix{\psi^{}_1\cr \psi_2^{}\cr}\ ,\ee
where $\psi^{}_{1,2}$ are two-component spinors which satisfy
\bea
(\sigma_1 T_{}^4+\sigma_2 T_{}^5+
\sigma_3 T_{}^6+iT_{}^7)\psi^{}_1&=&0\ ,\cr
(\sigma_1 T_{}^4+\sigma_2 T_{}^5+
\sigma_3 T_{}^6-iT_{}^7)\psi^{}_2&=&0\ .
\eea
These equations are most easily worked out in the Schwinger representation
of the $SU(3)$ generators

\be T_{}^A={\bf a}^\dagger{\lambda_{}^A\over 2}{\bf a}\ ,\ee
where the $\lambda^A$ are the Gell-Mann matrices, and $\bf a$(${\bf
a}^\dagger$) represent three
ladder operators $a^{}_\alpha$($a^{\dagger}_\alpha$), with the standard commutation
relations
\be[a^{}_\alpha,a_\beta^\dagger]=\delta^{}_{\alpha\beta}\ .\ee
If we further split the components as
\be
\psi^{}_{1,2}=\pmatrix{\eta^+_{1,2}\cr \eta_{1,2}^-\cr}\ ,\ee
these equations become
\be
\pmatrix{a_3^\dagger a_2^{}&a_3^\dagger a_1^{}\cr a_1^\dagger
a_3^{}&-a_2^\dagger a_3^{}\cr}
\pmatrix{\eta^+_{2}\cr \eta_{2}^-\cr} =0\ ,\qquad
\pmatrix{a_2^\dagger a_3^{}&a_3^\dagger a_1^{}\cr a_1^\dagger
a_3^{}&-a_3^\dagger a_2^{}\cr}
\pmatrix{\eta^+_{1}\cr \eta_{1}^-\cr} =0\ ,\ee
For each representation of $SU(3)$, there are three solutions of the
Kostant-Dirac equations.
The simplest solutions are just constant entries times the oscillator
vacuum state. Under the
diagonal subgroup, they split into two singlets 
\be
\vert\xi>\equiv \pmatrix{1\cr 0\cr 0\cr 0}\vert 0>\ ,\qquad \pmatrix{0\cr
1\cr 0\cr 0}\vert 0>\ ,
\ee
and one doublet
\be
\pmatrix{0\cr 0\cr 1\cr 0}\vert 0>\ ,\qquad \pmatrix{0\cr 0\cr 0\cr 1}\vert
0>\ .
\ee
The second Euler triplet solutions are made up of states with one oscillator.
They consist of one triplet with $L_8=1/(2\sqrt{3})$
\be
\pmatrix{0\cr 0\cr a^\dagger_1\cr 0}\vert 0>\ ,~\pmatrix{0\cr 0\cr
~a^\dagger_2\cr -a^\dagger_1}
\vert 0>\ ,~
\pmatrix{0\cr 0\cr 0\cr a^\dagger_1}\vert 0>\ ,\ee
one doublet with $L_8=2/\sqrt{3}$
and one singlet with $L_8=-5/(2\sqrt{3})$
\be
 \pmatrix{a^\dagger_1\cr 0\cr 0\cr 0}\vert 0>\ ,\pmatrix{a^\dagger_2\cr
0\cr 0\cr 0}\vert 0>
\ ;~~~~
\pmatrix{0\cr a^\dagger_3\cr 0\cr 0}\vert 0>\ .\ee
The ground state Euler multiplet can be generated by two nilpotent operators
$Q^{}_{\alpha}$, with $T^{}_8=-\sqrt{3}
/2$
\be
Q^{}_1=\pmatrix{0&0&0&0\cr
0&0&0&1\cr
1&0&0&0\cr
0&0&0&0\cr}\ ,\qquad
Q^{}_2=\pmatrix{0&0&0&0\cr
0&0&-1&0\cr
0&0&0&0\cr
1&0&0&0\cr}\ .\ee
They anticommute
\be
\{Q^{}_\alpha,Q^{}_\beta\}=0\ ,\ee
and satisfy
\be
[T^{}_i,Q^{}_{\alpha}]={i\over 2}(\sigma_i)_\alpha^\beta Q_\beta\ ,\qquad
[T^{}_8,Q^{}_{\alpha}]
=-i{\sqrt{3}\over 2}Q^{}_{\alpha}\ ,\ee
so that the ground states can be written in the usual supersymmetric
notation as
\be
\vert \xi>\ ,\qquad Q^{}_{1}\vert \xi>\ ,\qquad Q^{}_{2}\vert \xi>\ ,\qquad
Q^{}_1Q^{}_2\vert \xi>\ .\ee
It does not seem possible to write the states at the next level in the same
way; yet they carry much
of the same structure as those of the ground state.

One can build ladder operators which make transitions between one triple
and the next. There are
two such operators, a doublet $D_\alpha$ with $T_8^{}=1/(2\sqrt{3})$, and a
singlet $S$ with
$T^{}_8=-1/\sqrt{3}$, with matrix representation

\be D_{1,2}^{}=\pmatrix{a^\dagger_{1,2}&0&0&0\cr
0&0&0&0\cr
0&0&a^\dagger_{1,2}&0\cr
0&0&0&a^\dagger_{1,2}\cr}\ ,\qquad S=\pmatrix{0&0&0&0\cr
0&a^\dagger_3&0&0\cr
0&0&0&0\cr
0&0&0&0\cr}\ .\ee
Let us  normalize the value of $T_8^{}$ so as to give the singlets in
the lowest triple values of $\pm 1/2$ which are natural if one wants to
think of the $U(1)$ as
the Poincar\' e group
helicity. Then the ladder operators have $``$helicities" of $1/6$ and
$-1/3$, respectively, which
makes little sense, unless we consider the continuous spin representations
of the Poincar\' e
group. It is therefore doubtful that these operators can be explained in
the usual
 relativistic framework.

\section{Conclusions}
In this paper, we have drawn attention to certain intriguing facts:
\begin{itemize}
\item The massless degrees of freedom of certain local theories, some
of which are local limits of more elaborate theories (M-theory, type
IIB Superstring theory), can be viewed as solutions of an algebraic
Dirac-like equation. This raises the possibility of an alternate
formulation for these theories.

\item This Dirac-like equation can be identified with
the zero-mode of the superconformal generator in the Kazama-Suzuki
coset construction. These might therefore stem from 
theories with an underlying two-dimensional structure. 
\end{itemize}
Many questions remain unanswered, notably, 
\begin{itemize}
\item How is the ground state Euler multiplet selected?
\item What is the role of the other Euler multiplets?
\item  What is the spectrum of these superconformal theories?
\item  Is their spectrum  relativistic, and under what conditions? 
\item  What are the interactions?
\end{itemize}
Until these questions are answered, one cannot claim that the
suggestions in this paper are  physically relevant, but our list includes
many theories of great interest, namely $N=4$ 
super-Yang-Mills, type IIB superstring theory, and most notably, 
$N=1$ supergravity in eleven dimensions. The latter is the local limit
of M-theory, which raises the possibility that M-theory also 
displays  an underlying
two-dimensional structure, (perhaps holographically related to  a
membrane?). Equally intriguing is the appearance of the exceptional 
group $F_4$ in the description of its {\em space} degrees of freedom
of this theory.

{\em We dedicate this work to the memory of  the late Professor Golfand.}
\section*{Acknowledgements}
Lars Brink is supported in part by The Swedish Natural Research Council, while
Pierre Ramond is supported in part
by the US Department of Energy under grant DE-FG02-97ER41029.
Both have benefited from discussions with Professors B. Kostant and
S. Sternberg. One of us (PR) wishes to acknowledge the kind hospitality of  the Particle
Theory Group at the Enrico Fermi Institute and the Aspen Center for
Physics where some of this work was formulated, as well as  Professor N.
Warner for an illuminating discussion on superconformal algebras, and Mr.  T.
Pengpan for his help in working out some of the coset spectra.

\end{document}